\documentclass[twocolumn]{article}
\usepackage[utf8]{inputenc}
\usepackage{hyperref}
\usepackage{graphicx}
\usepackage{xcolor}
\usepackage{siunitx} 
\usepackage{amssymb}
\usepackage{amsmath}
\usepackage{ulem,lipsum}
\usepackage[a4paper,width=185mm]{geometry}

\begin{document}

\title{Majorana-like Coulomb spectroscopy in the absence of zero bias peaks}

\author{Marco Valentini$^{1,*,\dagger}$, Maksim Borovkov$^{1,5,\dagger}$, Elsa Prada$^2$, Sara Mart\'i-S\'anchez$^{3}$,  Marc Botifoll$^{3}$, \\ Andrea Hofmann$^{1,6}$, Jordi Arbiol$^{3,4}$, Ram\'on Aguado$^2$, Pablo San-Jose$^{2,*}$, Georgios Katsaros$^{1,*}$}

\date{%
    $^1$Institute of Science and Technology Austria, Am Campus 1, 3400 Klosterneuburg, Austria.\\
    $^2$Instituto de Ciencia de Materiales de Madrid (ICMM), Consejo Superior de Investigaciones Cient\'{i}ficas (CSIC), Sor Juana In\'{e}s de la Cruz 3, 28049 Madrid, Spain.\\ %Research Platform on Quantum Technologies (CSIC).\\
    $^3$ Catalan Institute of Nanoscience and Nanotechnology (ICN2), CSIC and BIST, Campus UAB, Bellaterra, Barcelona, Catalonia, Spain.\\
    $^4$ ICREA, Passeig de Llu\'is Companys 23, 08010 Barcelona, Catalonia, Spain.\\
    $^5$ Department of Physics, Princeton University, Princeton, NJ 08544. \\
    Present address:  $^6$ Universität Basel, Klingelbergstrasse 82, CH-4056 Basel. \\
    $\dagger$ These authors contributed equally to the work. \\
    $^*$ Corresponding authors: marco.valentini@ist.ac.at, pablo.sanjose@csic.es, georgios.katsaros@ist.ac.at.\\[2ex]%
    \today
}

\maketitle

\section{Abstract}
\textbf{%Hybrid semiconductor-superconductor devices hold great promise for realizing topological superconductivity. The edge states of this novel phase of matter are the long sought-after Majorana zero modes. 
Hybrid semiconductor-superconductor devices hold great promise for realizing topological quantum computing with Majorana zero modes~\cite{Nayak2008, Beenakker2013,Aasen2016,Aguado2017,Karzig:PRB17,Lutchyn2018,Prada:NRP20}. However, multiple claims of Majorana detection, based on either tunneling~\cite{mourik_signatures_2012,Das2012, Deng:S16,Nichele:PRL17,Vaitiekenaseaav3392} or Coulomb spectroscopy~\cite{albrecht_exponential_2016,van2016conductance}, remain disputed. %\sout{due to ambiguities of interpretation}. 
Here we devise an experimental protocol that allows to perform both types of measurements on the same hybrid island by adjusting its charging energy via tunable junctions to the normal leads. This method reduces ambiguities of Majorana detections by checking the consistency between Coulomb spectroscopy and zero bias peaks in non-blockaded transport. %Our experimental protocol greatly removes ambiguities, as it allows us to verify the consistency of a Majorana interpretation based on Coulomb-blockaded measurements (transition from Cooper pair to single electron peaks in conductance) concurrently with length-dependent zero bias peaks in non-blockaded transport. We find no such concurrent measurements hence ruling out topological Majorana modes.
 %This method allows us to rule out apparent topological Majorana modes \editP{in our devices}.
Specifically we observe junction-dependent, even-odd modulated, single-electron Coulomb peaks without concomitant low-bias peaks in tunneling spectroscopy. 
We provide a theoretical interpretation of the experimental observations in terms of low-energy, longitudinally-confined, island states rather than overlapping Majorana modes. 
Our method highlights the importance of combined measurements on the same device for the identification of topological Majorana bound states.}%Our  method investigates the consistency of the Majorana interpretation based on Coulomb-blockaded measurements (transition from Cooper pair to single electron peaks in conductance) with the appearance of zero bias peaks in non-blockaded transport. We find no such concurrent measurements hence ruling out topological Majorana modes. 
% Overall, our results demonstrate  that transport spectroscopy in both regimes strongly depend on how the superconducting island is connected to leads by tunneling junctions, hence highlighting the necessity of combined measurements in order to make progress in unambiguous detection of Majoranas in hybrid devices.

\begin{figure*}
\center{\includegraphics[]{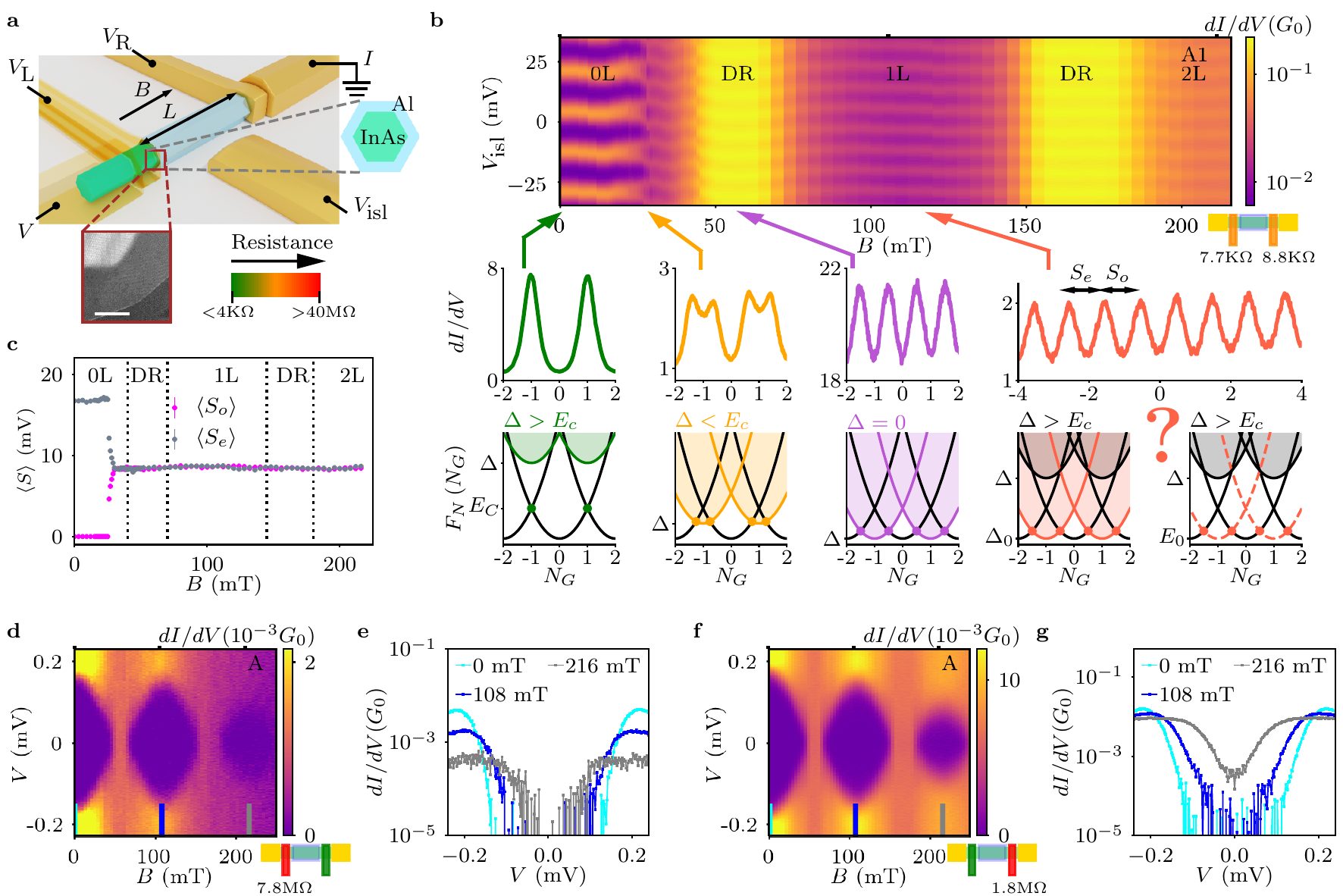}}
\caption{\textbf{Experimental protocol for combining tunneling and Coulomb spectroscopy on the same device.} (\textbf{a}) Schematics of a full-shell NW device. Turquoise represents the hexagonal InAs core, light blue the Al full-shell and gold the Ti/Au leads and gates. The AlOx layer separating the top gates and the tunnel junctions is not depicted. The insets show the NW cross-section schematic (right) and atomic resolution HAADF STEM image (left). The white scale bar corresponds to $20\si{nm}$.  (\textbf{b}) Zero-bias conductance $dI/dV$ plotted in logarithmic scale, in units of $G_0=2e^2/h$, as a function of $V_{\textrm{isl}}$ and magnetic field $B$ for device A ($L\approx 700$nm) with the tunnel barriers tuned to the intermediate-coupling regime. The ticks positioned on the upper axis of the figures correspond to integer multiples of the flux quantum $\Phi_0$. The NW free energy spectrum (see Eq. 1) and zero bias conductance  versus normalized gate voltage $N_G$ are shown below for $B = 0 \ \si{mT}$ (green), $B = 27 \ \si{mT}$ (orange), $B = 56 \ \si{mT}$ (violet) and $B = 116 \ \si{mT}$ (red). The conductance ranges are given in units of $10^{-2}G_0$. The 2-e to 1-e transport transition occurs within the Little-Parks zeroth lobe (0L) and 1-e-periodic spacing remains for the destructive regimes (DRs) and higher-order lobes (1L, 2L). (\textbf{c}) Coulomb peak spacing, $\left<S_e\right>$ and $\left<S_o\right>$, extracted from \textbf{b}. (\textbf{d}) [(\textbf{f})] $dI/dV$ as a function of $V$ and $B$ in the configuration where the left [right] junction is tuned to the weak coupling regime and the right [left] to the open regime. No signatures of sub-gap states are observed. (\textbf{e}) [(\textbf{g})] Line-traces taken from (\textbf{d}) [(\textbf{f})]  at the center of the 0L (light blue), 1L (dark blue) and 2L (gray), showing a hard gap in the 0L and no ZBPs in the 1L.} 
\label{fig:figure_1}
\end{figure*}

%Our theoretical analysis suggests that the single electron conductance peaks are the result of confinement taking place in the superconducting island rather than Majorana zero modes, highlighting therefore the necessity of combined measurements in order to unambiguously demonstrate the existence of the latter.} 

%\editP{To date, multiple claims of Majorana detection have been made based on different techniques, such as tunneling or Coulomb spectroscopy, applied individually. Here we devise an experimental protocol that allows for a combined application of these two techniques to the same device, greatly reducing the interpretation ambiguities of each of them individually. Our results challenge a number of past reports. We find that uniformly}

\section{Main}
One of the most hunted-for signatures of a Majorana zero mode (MZM) at the edge of a one-dimensional topological superconductor is a resonant zero bias peak (ZBP) in tunneling spectroscopy experiments \cite{lutchyn_majorana_2010,oreg_helical_2010,Lutchyn:NRM18,Prada:NRP20}. For hybrid nanowires (NWs) this was first reported in 2012 \cite{mourik_signatures_2012} with several other works appearing in the following years \cite{Das2012, Deng:S16,Nichele:PRL17,Vaitiekenaseaav3392}. However, by now it has become clear that peaks pinned to zero bias cannot be easily distinguished from those of non-topological origin. Unintentional quantum dots (QDs) \cite{Lee:NN14,ramon2015, marco_science_2021}, NW smooth confinement \cite{Prada2012,Kells2012, liu2017andreev,Fleckenstein:PRB18,Penaranda:PRB18,PhysRevB.98.245407,Moore2018,Vuik2019} and/or disorder in the semiconductor \cite{Chen2019,Ahn:PRM21} can give rise to topologically trivial zero energy states that mimic many of the properties of their topological counterparts. 
An alternative route to detect MZM signatures is Coulomb blockade spectroscopy \cite{albrecht_exponential_2016,van2016conductance}, where the transition from coherent two-electron (2-e) to one-electron (1-e) transport and the exponential length dependence of the 1-e Coulomb peak spacing are associated with the appearance of topological excitations. In contrast to tunneling spectroscopy measurements, the observation of parity transitions in Coulomb blockade appears surprisingly robust against electrostatic tuning \cite{carrad2020shadow,shen2018parity,shen2020full} and is ubiquitous for various material platforms \cite{pendharkar2021parity,kanne2021epitaxial,whiticar2020coherent,het2020plane}.  
%In order to relate the results from both spectroscopy techniques 
%In this work we devise a method based on the tuning of junction barriers that allows to compare the results of both spectroscopy techniques by applying them to  the same device. 
%Full-shell NWs, i.e., wires covered all around by the parent superconductor, are ideal for this purpose because MZMs are predicted to appear at well defined magnetic fields~\cite{Vaitiekenaseaav3392,fern2019evenodd}. In addition, the superconducting full-shell largely screens the electrochemical potential of the semiconductor core from external electric fields, so that the influence of junction tuning on a potential topological phase is minimized. 
%; therefore it is possible to tune the gates without affecting the NW electronic states.\newline

\section{Experimental Protocol}

In this work we devise a method based on the tuning of junction barriers that allows to compare the results of both spectroscopy techniques by applying them to  the same device. 
Full-shell NWs, i.e., wires covered all around by the parent superconductor, are ideal for this purpose because MZMs are predicted to appear at well defined magnetic fields~\cite{Vaitiekenaseaav3392,fern2019evenodd}. In addition, the superconducting full-shell largely screens the electrochemical potential of the semiconductor core from external electric fields, so that the influence of junction tuning on a potential topological phase is minimized. The scheme of our devices is depicted in Fig~\ref{fig:figure_1}a. We use full-shell Al/InAs NWs %\EditG{\sout{(wires covered on all facets by the parent superconductor)}} 
\cite{krogstrup_2015} and NWs for which the Al shell has been partially etched away (dubbed partial shell devices), in order to demonstrate the breadth of our results. The atomic resolution aberration corrected (AC) high angle annular dark field (HAADF) scanning transmission electron microscopy (STEM) images indicate a perfect epitaxy of the corresponding crystal lattices (see inset of Fig~\ref{fig:figure_1}a and Fig S1 in the supplementary information (SI)). A hybridized semiconductor-superconductor island of length $L$ is formed (see methods for fabrication details) and the transparencies of left and right junctions, consisting of bare InAs NW segments, are independently controlled by applying voltages $V_{L}$ and $V_{R}$ to the corresponding gates. The charge occupancy of the island is varied through the island gate voltage $V_{\textrm{isl}}$. The gate configuration allows an in-situ tuning of the barrier transparencies, which in turn control the Coulomb charging energy $E_C = \frac{e^2}{2 C_\mathrm{tot}}$ (where $C_\mathrm{tot}$ is the total capacitance of the island)~\cite{flensberg1994capacitance}. For sufficient negative gate voltages it is possible to deplete the bare InAs junction and therefore switch off the devices from the left and right side independently. The transparency of the left (right) junction is characterized by the device resistance when the right (left) junction is completely on (see section 2 in the SI).

We distinguish between the island regime and the open regime. The latter has at least one very transparent barrier (colored in green) and a conductance measurement versus bias voltage $V$ and $V_{\textrm{isl}}$ does not reveal any $E_C$. For the island regime, we differentiate between the weak coupling regime (barriers colored in red) and the intermediate-coupling regime (barriers colored in orange). In the weak tunneling regime the subgap conductance at zero field is below the experimental noise level, while in the intermediate spectroscopy regime increased subgap conductance is measured due to Andreev tunneling \cite{PhysRevB.25.4515}.  

%Approximately 100 devices were pre-characterized at mK temperatures, from which we chose to investigate systematically only those that exhibited tunable junctions. We moreover perform $dI/dV$ measurements as a function of $V$ and $V_{L(R)}$ at zero field and exclude devices that show Andreev bound states in the junctions~\cite{marco_science_2021} (Fig. S3). %Also devices indicating quasiparticle poisoning are not considered in this study. 
Here we report the findings for 7 full-shell islands ranging from 200 nm to 1 $\si{\mu m}$, which we refer to as A ($L \approx 700 \si{n m}$), B ($L \approx 400 \si{n m}$), C ($L \approx 900 \si{n m}$), D ($L \approx 200 \si{n m}$), E ($L \approx 600 \si{n m}$), F ($L \approx 950 \si{n m}$) and G ($L \approx 1 \si{\mu m}$) and 2 partial shell islands which we refer to as pA ($L \approx 600 \si{n m}$) and pB ($L \approx 550 \si{n m}$). 

We first focus on the Coulomb spectroscopy of device A (700 nm), see Fig.~\ref{fig:figure_1}b. %With the two tunnel barriers tuned to the strong coupling regime, 
We investigate the Coulomb peaks in zero-bias differential conductance as a function of island gate $V_{\textrm{isl}}$ and parallel magnetic field $B$. The magnetic flux along the superconducting shell gives rise to the Little-Parks (LP) effect, whereby the superconducting gap $\Delta(B)$ is periodically modulated by the flux $\Phi$ penetrating the NW cross section \cite{LP_1962,PhysRevB.101.060507}. The magnetic field intervals for which $\Delta \neq 0$, called as zeroth (0L), first (1L) and second lobe (2L), alternate with intervals where $\Delta = 0$, dubbed as destructive regimes (DRs). Each $n=0,1,2...$ lobe is associated with a number $n$ of twists (or \textit{fluxoid number}) of the superconducting phase around the shell, that grows by one each time the magnetic flux increases by a flux quantum $\Phi_0=h/2e$.
%, every time that another flux quantum $\Phi_0=h/2e$ threads the superconducting shell.

To qualitatively understand the Coulomb blockade spectroscopy, we invoke the phenomenological energetic model \cite{Tuominen1992PRL,andrew_nature_phys}. In thermodynamic equilibrium, the Gibbs free energy of an island with charge occupancy $N$ and dimensionless gate-induced charge $N_G = \eta V_{\textrm{isl}} / E_C$, is given by:
\begin{equation}
    F_N(N_G, B) = E_C (N - N_G)^2 + \Delta(B) N \mathrm{mod} 2,
\end{equation}
where $\eta$ is the lever arm. We perform the Coulomb spectroscopy measurement in the regime where $E_C < \Delta(B=0)$. For each $N$, odd or even, $F_N$ has a parabolic shape when changing $N_G$ and the degeneracy points between two parabolas lead to Coulomb peaks in zero-bias conductance (marked as filled circles in the bottom insets of Fig.~\ref{fig:figure_1}b). For zero magnetic field, Andreev processes give rise to coherent 2-e transport at these free energy degeneracy points as one electron drags a second one to create a Cooper pair \cite{hekking_prl_1993} (green inset in Fig.~\ref{fig:figure_1}b). Upon increasing the field, $\Delta(B)$ decreases and eventually vanishes, bringing the odd charge occupancy parabolas to zero energy. The 2-e peaks split at $B = 27 \si{mT}$ (orange inset in Fig.~\ref{fig:figure_1}b) as soon as $E_C > \Delta(B)$. When $\Delta(B)=0$ (DR), the island exhibits 1-e periodicity signaling metallic single electron transport (violet inset in Fig.~\ref{fig:figure_1}b).

\begin{figure} [h!]
\center{\includegraphics[]{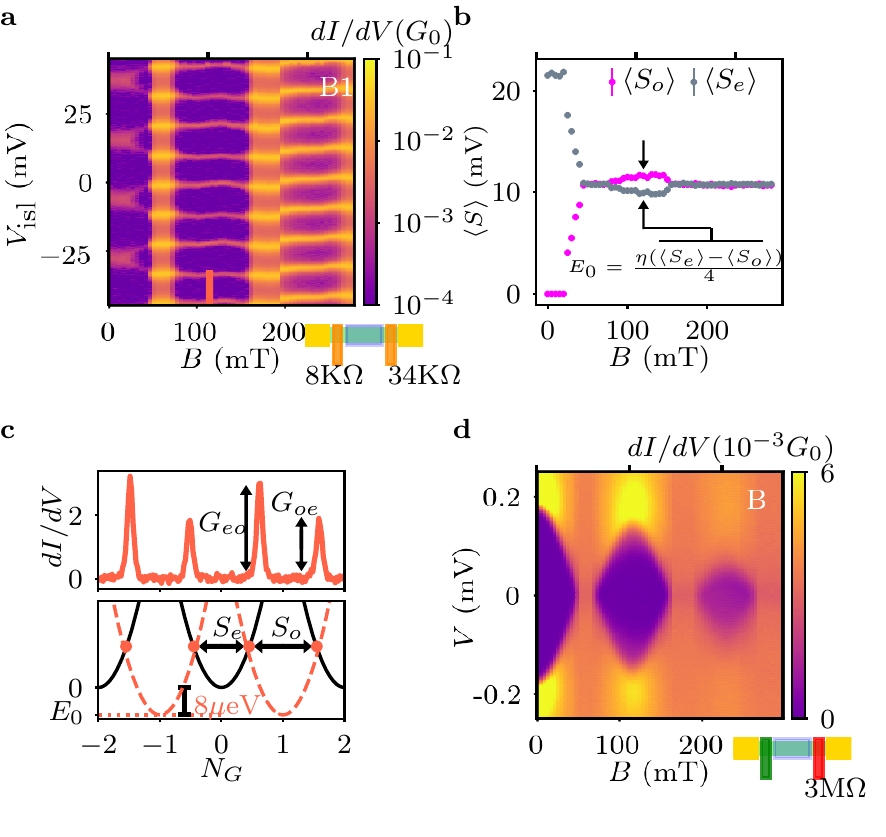}}
\caption{\textbf{Coulomb and tunneling spectroscopy for a device with even-odd modulation.}
(\textbf{a})  Zero bias $dI/dV$ as a function of $V_{\textrm{isl}}$ and $B$ in the intermediate-coupling regime. (\textbf{b}) Coulomb peak spacing, $\left<S_e\right>$ and $\left<S_o\right>$, extracted from \textbf{a}.
(\textbf{c}) (top) Line-trace of \textbf{a} taken at $B = 115 $mT where the Coulomb peaks are not evenly spaced, i.e. $\left<S_e\right> \neq \left<S_o\right>$ (in units of $10^{-3} G_0$). Consecutive peak heights are denoted by $G_{eo}$ and $G_{oe}$. (bottom) Energy of the system obtained from the panel above. The peak spacing in the upper panel corresponds to having a lowest energy state $E_0 \approx 8 \mu$eV, see SI for details. (\textbf{d}) $dI/dV$ as a function of $V$ and $B$ with $V_{R}$ tuned in the weak tunneling regime and $V_{\textrm{L}}$ tuned in the open regime, performing therefore tunneling spectroscopy from the right side.} 
\label{fig:figure_2}
\end{figure}

As we increase the field further, the superconducting LP gap reopens and the condition $E_C < \Delta(B)$ for 2-e transport should be restored. However, instead of 2-e, we observe 1-e periodic transport (red inset in Fig.~\ref{fig:figure_1}b), similar to the DR transport. %Just the background conductance between the Coulomb peaks is slightly higher than that measured in the DR. 
As a spectroscopic figure of merit, we compute the average spacing for odd and even Coulomb peaks, $\left<S_o\right>$ and $\left<S_e\right>$, corresponding to odd or even ground states, respectively. Fig.~\ref{fig:figure_1}c depicts $\left<S_o\right>$ and $\left<S_e\right>$ as a function of $B$ extracted from Fig.~\ref{fig:figure_1}b. No measurable difference between $\left<S_o\right>$ and $\left<S_e\right>$ is observed in the DRs, in the 1L or in the 2L (Fig.~\ref{fig:figure_1}c).

The 1-e transport could be explained either with a gapless superconducting island or with a subgap level of energy $E_0\approx 0$, such as a MZM, see red inset of Fig.~\ref{fig:figure_1}b. 
In order to clarify the validity of either of these scenarios we perform tunneling spectroscopy measurements at both ends of device A. It is important to note that there is one major difference between the two types of measurements. For Coulomb spectroscopy we need to form an island with $E_C$, while for tunneling spectroscopy we need to tune one of the junctions to the open regime in order to suppress $E_C$.
In Fig.~\ref{fig:figure_1}d we tune $V_{R}$ to the open regime and $V_{L}$ to the weak coupling regime and we investigate the evolution of $dI/dV$, as a function of $B$ and $V$. Surprisingly, we cannot confirm at first sight
either of the two possible scenarios. We observe no measurable conductance within a hard gap that follows the LP modulation, and no trace of a ZBP in the 1L consistent with an $E_0\approx 0$ state that could support 1-e periodic transport in CB (Fig.~\ref{fig:figure_1}e). %Similarly we perform tunneling spectroscopy at the right side, see Fig.~\ref{fig:figure_1}e. 
We then reverse the voltages at the top gates, closing $V_{R}$ and opening $V_{L}$ and perform tunneling spectroscopy at the right side (see Fig.~\ref{fig:figure_1}f). We again note the LP modulation and no subgap conductance above the noise level of our measurement or ZBP, as highlighted by the line-traces in Fig.~\ref{fig:figure_1}g. \newline
In order to understand whether the NW length has an effect on the above observations, we investigate device B which has a shorter length, $L \approx 400 \ \si{nm}$, and perform again tunneling and Coulomb spectroscopy.
Figure ~\ref{fig:figure_2}a shows the Coulomb spectroscopy measurement with the tunnel barriers in the intermediate-coupling regime. Like in the previous device, the 0L is dominated by 2-e coherent transport. 
Again, in the 1L the 2-e coherent transport is not recovered. Unlike the previous case where Coulomb peaks where equally spaced in the 1L, see Fig.~\ref{fig:figure_1}c, here the Coulomb peaks exhibit an even-odd pattern ($\left<S_e\right> \neq \left<S_o\right>$), see Figs.~\ref{fig:figure_2}b and c.   From $\left<S_e\right>$ and $\left<S_o\right>$ it is possible to extract the energy scale $E_0 = \eta \left(  \frac{ \langle S_e \rangle - \langle S_o \rangle}{4} \right)$. In the limit of low temperatures and contact transparencies, $E_0$ can be interpreted as the lowest electron-like subgap level (see SI for details) \cite{PhysRevB.97.041411}, although this connection should be expected to break down at higher transparencies, where higher-order transport processes dominate. At the middle of the 1L we can extract an energy of $E_0\approx -8 \mu$eV (Fig.~\ref{fig:figure_2}c bottom), i.e., very close to zero energy. Due to finite temperature in the experiment, a subgap level with such a small energy scale would lead to a broadened but visible ZBP in tunneling spectroscopy, since the $3.5k_{B}T$ Fermi broadening is about 15 $\mu$eV in our experiment (the effective electron temperature is about 50mK). However, the tunneling spectroscopy measurements in the weak tunneling regime from both island sides do not reveal any ZBP or other discrete subgap states in the first lobe (Fig.~\ref{fig:figure_2}d and SI).

\begin{figure}[h!]
\center{\includegraphics[scale=1]{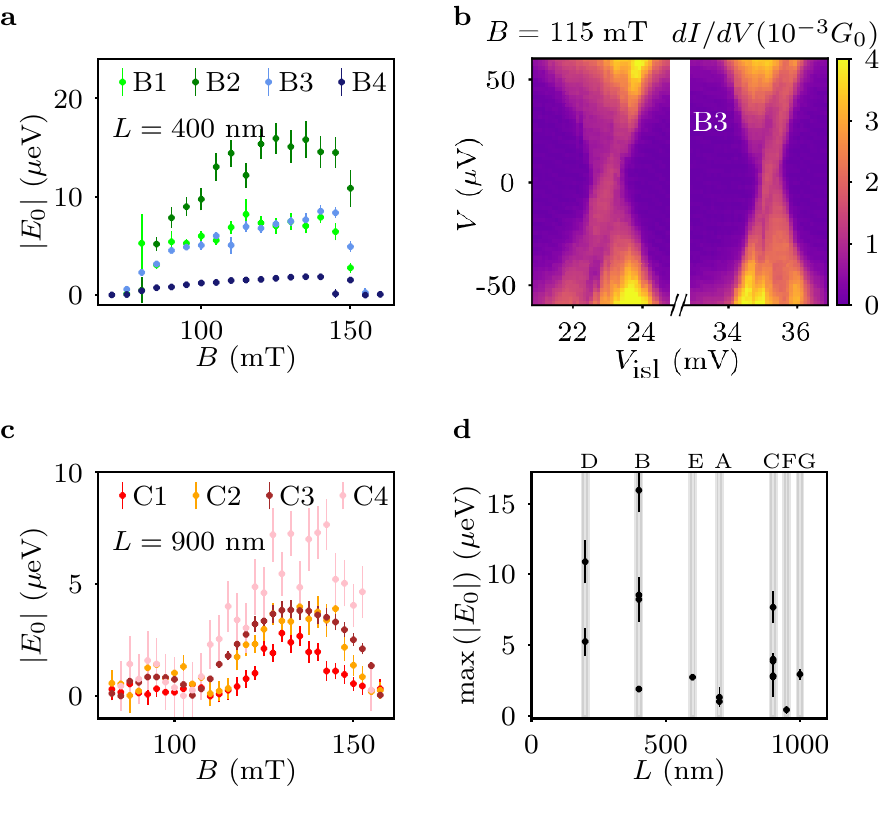}}
    \caption{\textbf{ Tunability of even-odd modulation by junction gate voltages for different island lengths.} (\textbf{a}) $|E_0|$ vs. $B$ in the 1L for 4 different configurations of device B, which has a top gate and  $L \approx 400 \ \si{nm}$. (\textbf{b}) $dI/dV$ as a function of $V$ and $V_{\textrm{isl}}$ at $B = 115 \ \si{mT}$, for configuration B3, demonstrating the presence of a ZBP and excited states. (\textbf{c}) $|E_0|$ vs. $B$ in the 1L for 4 different configurations of device C, which has a side gate and $L \approx 900 \ \si{nm}$.  (\textbf{d}) $\mathrm{max} \{ |E_0| \}$ as a function of $L$. The grey columns include the different configurations of each device. It is clear that $E_0$ depends, at least equally important, on the junction transparencies as on the island length. It is important to notice that in previous works \cite{albrecht_exponential_2016,Vaitiekenaseaav3392}, the reported value is $A = 4 E_0$. For Majoranas $E_0$ should follow an exponential-decaying law with $L$.}
\label{fig:figure_3}
\end{figure}

\section{Controlling the even-odd asymmetry }

To further investigate the nature of the even-odd modulation of the Coulomb peaks, we perform the same Coulomb spectroscopy experiment for different junction resistances of device B. In Fig.~\ref{fig:figure_3}\textbf{a} we plot the absolute value of $E_0$ as a function of $B$ taken in the 1L for four different junction configurations (B1, B2, B3 and B4; see Table~ST3 for details about the junction resistances). Clearly, the voltages applied to the junctions strongly influence the $E_0$ modulation, independently of island length $L$. For instance, at $B = 120\ \si{mT}$, $E_0$ can be tuned from above $15.4 \ \si{\mu eV}$ (configuration B2) to $1.6  \ \si{\mu eV}$ (configuration B4). Figure \ref{fig:figure_3}\textbf{b} furthermore demonstrates that transport takes place through a discrete resonance at low $V$, which is separated from a continuum at higher bias. However, we note that the observation or not of such a ZBP depends on the barrier configuration (see section 11 in the SI). In Fig. \ref{fig:figure_3}\textbf{c} we repeat the same analysis for device C, a long island device with $L \approx 900 \ \si{nm}$. Also for this device, it is possible to tune $E_0$ at $B = 140 \ \si{mT}$ from $7.3 \ \si{\mu eV}$ (configuration C4) to $2 \ \si{\mu eV}$ (configuration C1).
% These measurements demonstrate that any length dependence study of the CB even-odd modulation aiming to establish the presence of MZMs should be performed at exactly the same junction configurations for all lengths.

As a next step, we compare all the devices and all the configurations in terms of the maximum value of $E_0$ in the 1L. Fig.~\ref{fig:figure_3}\textbf{d} shows $\mathrm{max}\{ E_0\}$ as a function of $L$, with each grey column indicating the same device but different junction resistance configurations, see also Fig. S22. Once more, one observe that the junction resistances have an influence on $E_0$ comparable to the island length. A complementary figure of merit that could shed light on the potential presence of MZMs is the height difference between the odd Coulomb peak $G_{eo}$, and the even one $G_{oe}$, see top panel of Fig. \ref{fig:figure_2}\textbf{c}. Similarly, the peak height modulation strongly depends on junction properties (see SI).

\begin{figure*} 
\center{\includegraphics[width=\textwidth]{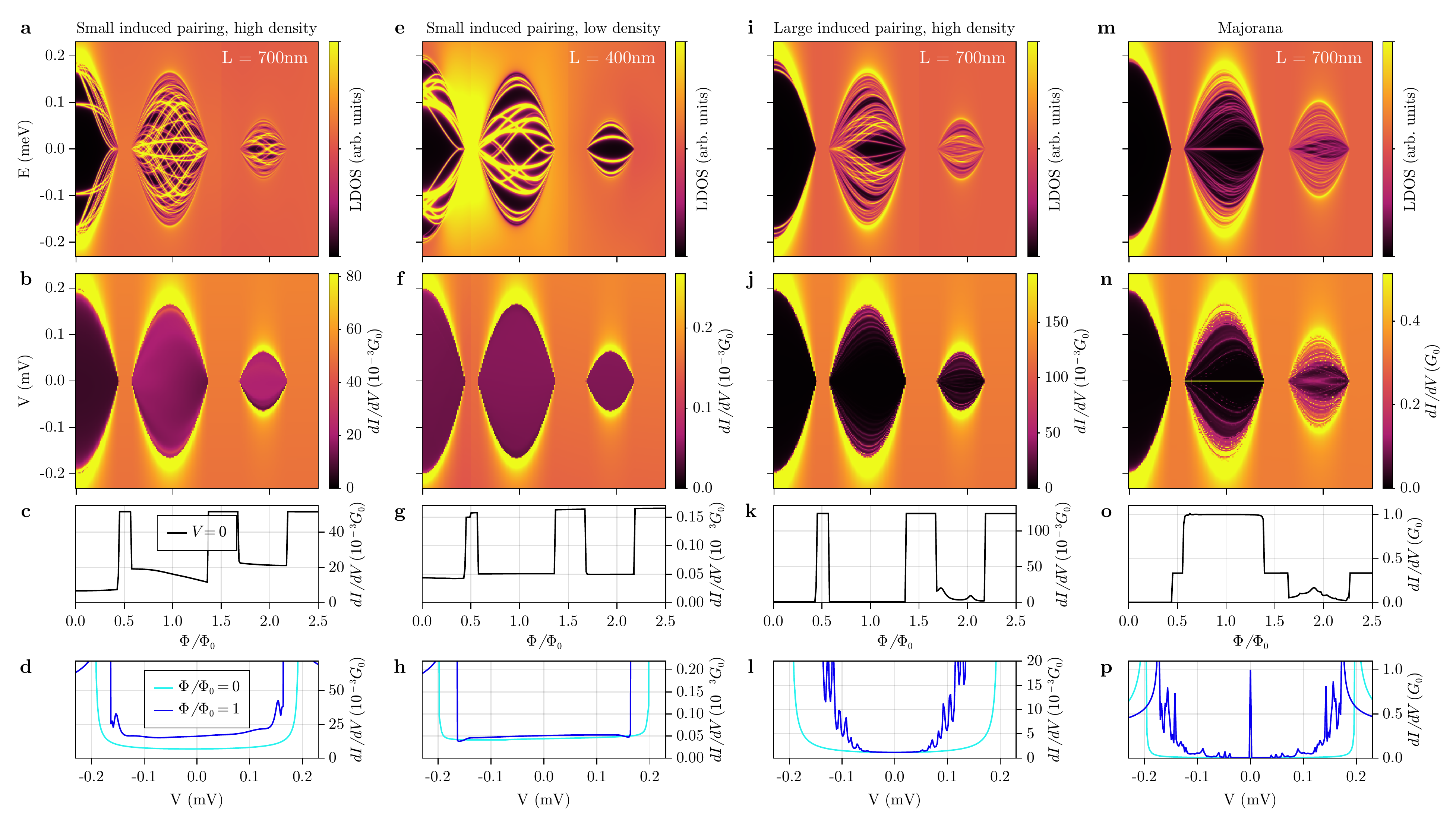}}
    \caption{\textbf{Numerical simulation of the single-particle local density of states (LDOS) and conductance ($dI/dV$) in full-shell nanowires.} (\textbf{a,e,i,m}) LDOS in arbitrary units as a function of normalized flux $\Phi/\Phi_0$ and energy $E$ for four different nanowire regimes: small induced pairing and high density, small induced pairing and density, large density and pairing and topological. The first two are the most representative of our experimental results. They differ only in the amount of charge transfer from shell to core and in the island length (see top-right corner), and exhibit a negligible and finite $E_0$, respectively. The third, like the second, has a finite $E_0$ but is rather insensitive to barrier configuration. The topological case is carefully tuned to have two zero energy Majorana bound states within the first lobe (spatially localized at the barriers) by increasing the spin-orbit coupling sufficiently and tuning the chemical potential. All are weakly coupled to normal reservoirs across 100 nm long tunnel barriers at either end. The LDOS is taken at a nanowire section adjacent to the left barrier. In the limit of low temperatures and high barrier resistance, the subgap state of lowest energy $E_0$ controls the 1e period modulation in CB spectroscopy. (\textbf{b,f,j,n}) Tunneling $dI/dV$ conductance as a function of bias $V$ and flux in the same four systems, where the right-barrier is now fully open, deconfining subgap states into the right (normal) lead. The subgap features in the LDOS of the $n=1$ lobe become broadened. This is visible in conductance either as a smooth uniform background in the first two cases (strong broadening), or as weak and sharper U-shaped subgap features for large pairing, due to residual normal reflection at the right contact, (\textbf{l}). In contrast, a strong zero-bias anomaly due to the localized MBS on the left barrier remains clearly visible in the Majorana case (\textbf{f}). \textbf{(c,g,k,o)} $V=0$ cuts of the corresponding $dI/dV$. \textbf{(d,h,l,p)} vertical cuts at $\Phi/\Phi_0=0,1$. All model parameters are given in Table ST14 of the SI.}
\label{fig:theory}
\end{figure*}

\section{Theory and discussion}

In this final section we develop a theoretical picture, a variation of the gapless superconductor scenario, that resolves the apparent conflict between transport and CB spectroscopy. We focus on the following key experimental observations: (1) the $dI/dV$ follows the LP effect but is otherwise rather featureless within the lobes;
(2) the measured CB conductance peak spacing corresponds to a $\sim 2$e charging periodicity in most of the $n=0$ lobe, but is otherwise $\sim 1$e-periodic, with an even-odd modulation in $n=1$ corresponding to an $|E_0|\approx 0-20\si{\mu eV}$ energy scale that depends strongly on sample and barrier configuration. 

Our model of the islands follows closely the formalism of Ref. \cite{Vaitiekenaseaav3392} (see Methods and SI). It allows us to simulate both the local density of states (LDOS) at any point of the island in a closed/closed barrier configuration (we choose a point inside the island adjacent to the left barrier), and also the $dI/dV$ transport for closed/open islands. 
The results are summarized in Fig. \ref{fig:theory}. We first compute the subgap spectrum of a closed/closed $L=700$nm-long island in a regime of high density and relatively small induced superconducting pairing on core states. We include spin-orbit coupling (SOC), and choose a g-factor $g=12$. At zero magnetic field the longitudinally quantized normal-state levels inside the core are pushed away from zero energy due to the shell-induced pairing. 
However, a finite magnetic flux quickly breaks the induced pairing, which leads to a dense quasicontinuum of energy levels crossing zero in $n>0$ lobes, and even close to the edges of the $n=0$ lobe. This is consistent with the kind of unmodulated 1e-periodic CB phenomenology of device A, see \ref{fig:figure_1}(b,c), which would follow from the gapless metallic-like subgap quasicontinuum in the closed/closed small pairing regime (so $E_0=0$, understood as the energy of the lowest non-interacting excitation).

As the right barrier is opened to perform tunneling spectroscopy, the quasicontinuum of states is deconfined into the normal lead, smearing out the LDOS in the $n>0$ lobes into a weak uniform background inside the gap. Its uniformity is the result of strongly suppressed normal reflection at the right contact, made possible by the smallness of the pairing and the low normal resistance of the right barrier. The weak, uniform LDOS becomes visible in our simulations as a small but finite increase with $n$ of the zero-temperature subgap $dI/dV$ conductance background, see Fig. \ref{fig:theory}(b-d).  
Note that this increase was not detected in the $dI/dV$ measurements of Figs. \ref{fig:figure_1} and \ref{fig:figure_2}, an apparent conflict that will be addressed later on.

The regime of high density and small pairing above explains the CB spectroscopy of samples with $E_0\approx 0$, like samples A or F. The emergence of a finite $E_0$ in other samples can be understood by considering a different regime with a shorter length $L$ and/or a lower electronic density. In this case, the $n\neq 0$ subgap states acquire a larger level spacing, see Fig. \ref{fig:theory}e. The lowest of these longitudinally confined states has now a finite energy $E_0\sim 1/L$, around $20 \mathrm{\mu eV}$ at its maximum in this instance. $E_0$ fluctuates strongly and stochastically with magnetic field and, notably, with barrier configuration, as the latter directly affects its longitudinal quantization condition, see Sec. 13.4 of the SI. Despite the complex LDOS dominated by mesoscopic fluctuations, and similar to the case of high density, opening the right barrier leads to a featureless $dI/dV$ with no visible trace of $E_0$, see Fig. \ref{fig:theory}(f-h). The increase in conductance background is much smaller in this case, but is still present.
A third plausible scenario, also with a finite $E_0$, is shown in Fig. \ref{fig:theory}(i-l). In this regime, both density and induced pairing are high, but $E_0$ exhibits a much smaller dependence of $E_0$ with barrier configuration, as it is dominated by the induced pairing, see SI. In summary, the range of behaviors of our model is consistent with the large variability of $E_0$ in our experiments. It also supports our observation that $E_0$ tends to decrease for longer islands, as also found in other experiments \cite{Vaitiekenaseaav3392}.

The above phenomenology is in stark contrast to the one expected in a topological island with Majorana bound states. The topological regime, as discussed in Ref. \cite{Vaitiekenaseaav3392}, is achieved by appropriate tuning of density and SOC. In Figs. \ref{fig:theory}(m-p) we have chosen a chemical potential in a topological phase and SOC has been artificially increased 25-fold, as required to suppress the Majorana splitting across the $L=700$nm island. The LDOS exhibits two Majorana bound states pinned to $E_0 = 0$. The left Majorana is localized close to the left barrier, so it is not significantly broadened upon opening the right barrier, unlike the right Majorana that readily delocalizes into the reservoir \cite{San-Jose:SR16,Avila2019}. The zero-temperature $dI/dV$ shows a strong zero-bias anomaly, of universal height $G_0=2e^2/h$ in the limit of zero temperature and splitting, in contrast to the non-topological regimes in Figs. \ref{fig:theory}(b,f). Note, however, that at finite temperatures and deep in the tunneling regime, the $G_0$ peak height can be strongly reduced, so reducing tunnel resistance may be required to observe it \cite{Setiawan2017}.

\begin{figure}[h]
\center{\includegraphics[scale=1]{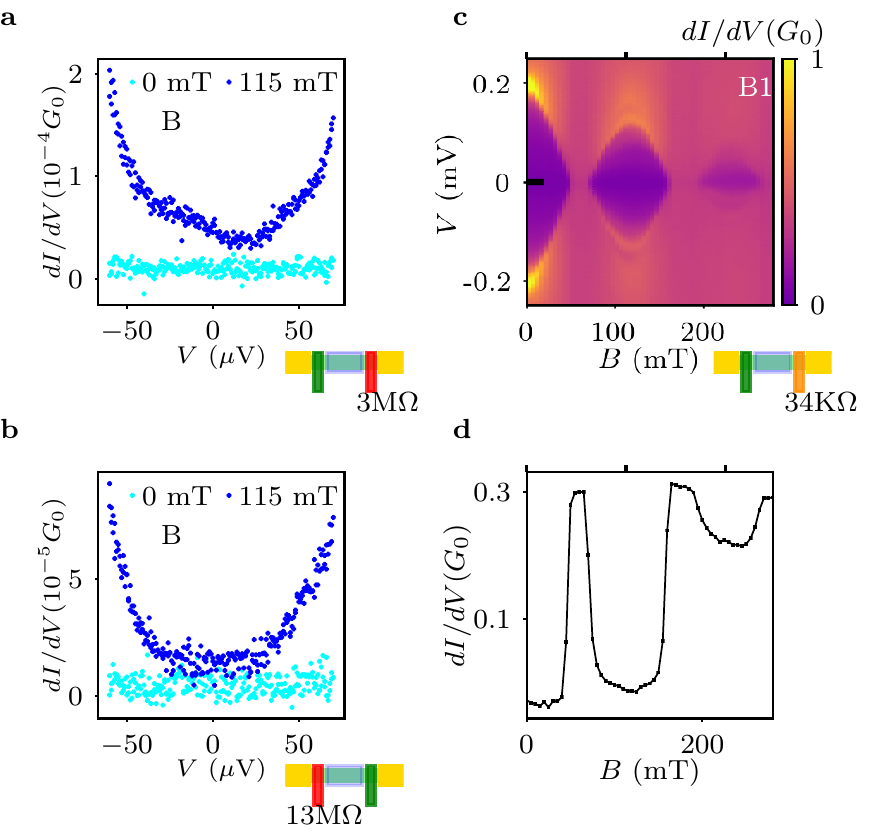}}
    \caption{ \textbf{Conductance in LP lobes when increasing the measurement sensitivity or barrier transparency.} (\textbf{a}) [(\textbf{b})] $dI/dV$ of device B as a function of $V$ taken in the middle of the 0L (dark blue) and the 1L (light blue) when performing the spectroscopy measurement from the left [right] side. It becomes obvious the subgap conductance in the 1L is enhanced. In order to resolve such low background conductance we decreased the noise floor at the expense of the total measurement time. (\textbf{c}) Intermediate-coupling spectroscopy performed from the right side. (\textbf{d}) Zero-bias conductance trace extracted from \textbf{d} showing how the conductance increases within the LP lobes. $dI/dV$ is plotted in logarithmic scale.}
\label{fig:gapless}
\end{figure}

\begin{figure}[h!]
\center{\includegraphics[]{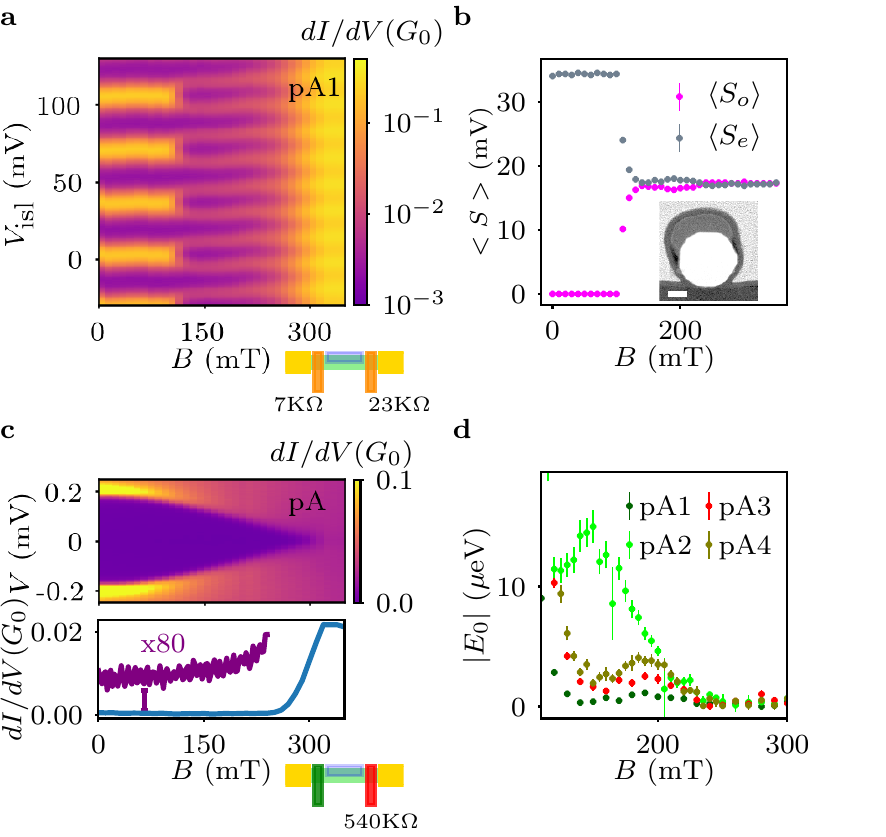}}
\caption{\textbf{Behaviour of partial-shell devices.} (\textbf{a}) Zero bias $dI/dV$ as a function of $V_{\textrm{isl}}$ and $B$ for configuration pA1. (\textbf{b}) Plot showing the Coulomb peak spacing extracted from \textbf{a}. In the inset, a TEM image of a partial-shell device is shown. The white area is the InAs core and the Al, shown in grey, does not cover all the facets but only the upper part of the wire. The scale bar corresponds to $20 \si{nm}$. (\textbf{c}) (top) $dI/dV$ as a function of $V$ and $B$ for device pA with one junction in the open regime and the other tuned in the weak coupling regime. The tunneling spectroscopy does not reveal a ZBP and/or subgap states. (bottom) Zero bias $dI/dV$ vs $B$. The purple curve in the inset shows that the conductance background grows with $B$. (\textbf{d})   Plot showing $E_0$ vs. $B$ for different configurations of the partial-shell device pA, i.e. pA1, pA2, PA3 and pA4. }
\label{fig:partial}
\end{figure}

The validity of our theoretical interpretation for the small or zero $E_0$ energy is contingent on the existence of a weak subgap background in the tunneling $dI/dV$ that increases in $n\neq 0$ lobes relative to $n=0$. This background increase stems from sequential tunneling through the LDOS quasicontinuum, that is ultimately responsible for the modulated 1-e periodicity of CB spectroscopy. (An $n$-independent cotuneling background is also expected for smaller $L$.)  Our initial measurements failed to reveal this background increase. To verify the consistency of our interpretation we revisit the measurements of $dI/dV$ versus $V$ by increasing the resolution and decreasing the noise floor. As shown in Fig. \ref{fig:gapless}(a,b), this allows us to detect a small but clear increase in subgap background conductance in the 1L as compared to the 0L (tunneling $dI/dV$ in device B measured from left and the right contacts, respectively), as expected from our theory. The effect is also visible in the intermediate-coupling regime, see Figs. \ref{fig:gapless}(c,d) and \ref{fig:theory}(b,c). Intermediate coupling softens the gap, but often also reveals subgap states similar to those in Fig. \ref{fig:theory}j, see SI. 
%The line trace at zero bias (Fig ...d) clearly demonstrates the increased subgap conductance as one moves from the 0L to the 1L and the 2L, in qualitative agreement with the theory traces (Fig. XX).
We note that the absence of a ZBP in the intermediate-coupling spectroscopy discards also the possibility that a MZM was not detected due to thermal broadening.

Finally, we investigate the generality of our findings by applying our experimental protocol to partial-shell NWs. %The TEM of such a partial shell device can be seen in Fig.~\ref{fig:partial}a. 
Fig.~\ref{fig:partial}a show the zero bias $dI / dV$  as a function of $V_{\textrm{isl}}$ and $B$ for device pA (configuration pA1). As expected, the LP effect does not take place, and the superconducting gap gradually decreases till around 300 mT, where $\Delta =0$. The zero bias transport is 2-e periodic from 0 mT to 110 mT. At this point the 2-e Coulomb peaks split and develop an even-odd pattern, which becomes 1-e periodic at around 300 mT where superconductivity is completely destroyed. 
Fig.~\ref{fig:partial}b depicts $\left<S_o\right>$ and $\left<S_e\right>$ as a function of $B$ extracted from Fig.~\ref{fig:partial}a. It is clear that $\left<S_e\right> \neq \left<S_o\right>$ from $B= 100 \ \si{mT}$ to $B= 230\ \si{mT}$, whereas for higher values of the field the transport becomes 1-e periodic. The extracted $E_0$ (see below) is below  $7.5 \ \si{\mu eV}$ and should be therefore observed as a ZBP.  
Similar to the full-shell devices, no ZBPs are observed in the tunneling spectroscopy measurement (see Fig.~\ref{fig:partial}c).  Similarly, no states appear in intermediate-coupling spectroscopy carried out with the same junction resistance as in Fig.~\ref{fig:partial}a, see Fig. S16. %This trend is typical also for the other partial-shell devices, see \EditG{Fig.~S XX} . 

These observations support the previous findings from the full-shell devices that the the origin of the observed parity oscillations are not MZMs, but longitudinally-confined states brought down to low energies by the magnetic field through orbital effects. The orbital mechanism was also studied in Ref. \cite{shen2020full} in the context of CB for partial-shell islands. For the sake of completeness, we studied the $E_0$ modulation for different barrier resistances also for the partial-shell devices. Fig.~\ref{fig:partial}d shows $E_0$ as a function of $B$ for 4 configurations of device pA. Again, $E_0$ can be dramatically tuned by the junction resistances. Device pB, which shows similar behavior, is discussed in the SI (Fig. S36).

In summary, we have shown that 1-e Coulomb peak spacing results from a quasi-continuous subgap spectrum at finite field. Junction-dependent longitudinal confinement may lead to even-odd CB spacing modulation without visible low bias peaks in tunneling spectroscopy. %\EditG{emphasizing that the application of both techniques on the same device allow to identify states as trivial}. 
Our work highlights the importance of junction details, even without quantum dot formation \cite{marco_science_2021}, for understanding transport spectroscopy in the quest for topological MZMs. Since topological edge states should theoretically be insensitive to such details, combined measurements on the same device like the ones presented here are an important step towards reducing ambiguities in interpretation. Length-dependence studies of MZM protection should take into account the possibility of junction-dependent longitudinally-confined modes as potential MZM false positives, and carefully discern algebraic ($1/L$) from exponential decays of even-odd modulations with length (see Fig.  S42)~\cite{Vaitiekenaseaav3392,albrecht_exponential_2016}.
% These measurements demonstrate that any length dependence study of the CB even-odd modulation aiming to establish the presence of MZMs should be performed at exactly the same junction configurations for all lengths.

\subsection*{Methods}

\subsubsection*{Fabrication}
We use full-shell NWs which offer advantages in the search of MZMs, as they are predicted to appear at well defined magnetic fields~\cite{Vaitiekenaseaav3392,fern2019evenodd}. In addition, the superconducting full-shell largely screens the electrochemical potential of the semiconductor core from external electric fields. The hexagonal InAs NWs, with a diameter of about $110 \, \si{nm}$, were grown via the vapor-liquid-solid (VLS) technique and the  $30 \, \si{nm}$ Al shell was epitaxially grown in situ, giving a highly transparent semiconductor-SC interface \cite{krogstrup_2015}. 
The NWs were deposited on a heavily doped silicon substrate covered with $285 \, \si{nm}$ of silicon oxide. The superconducting island is formed by selectively etching Al from both sides of the NW (for the partial shell devices Al is also removed from some NW facets.) The bare InAs segments at the edges of the island are used as tunnel barriers. In all devices the tunnel junction length was chosen to be below 120 nm, in order to avoid a QD formation in the tunnel barrier regions~\cite{marco_science_2021}. The island gates and contacts consist of normal-metal ($5 \, \si{nm}$/$180 \, \si{nm}$ Ti/Au bilayer). Two type of devices have been fabricated with top gates and with side gates. For the first type, a $\approx 9 \ \si{nm}$ thick layer of aluminium oxide is added at 150 $^{\circ}$C by plasma atomic layer deposition, followed by deposition of normal-metal top gates ($5 \, \si{nm}$/$180 \, \si{nm}$ Ti/Au bilayer). For the second type, the side gates are fabricated together with the ohmic contacts and the layout of the gates can be found in Fig.S27. %TEM studies do not reveal any change in the InAs/Al interface after device fabrication (see SI). 

\subsubsection*{Device sampling/Short junction effect}
 Approximately 110 devices showed current at mK temperature when applying a voltage to the source and measuring a current through the drain. The tunnel junctions of the realized devices were purposefully fabricated in the short-junction limit in order to suppress QD formation~\cite{marco_science_2021} (see SI for device details). However, 18 of those showed ABS in at least one of the junctions  and were therefore discarded. Furthermore, 78 did not show the required junction tunability, which we attribute to the short channel effect~\cite{short_junction_effect}. Finally 10 devices showed split Coulomb peaks and poisoning and were not further studied. The remaining 9 were fully characterized and are the focus of the present work.%nine had both junctions tunable by the gates. We attribute this low yield to the short channel effect~\cite{short_junction_effect}, as the tunnel junctions were purposefully fabricated in the short limit in order to \EditG{suppress} QD formation~\cite{marco_science_2021} (see SI for device details). %Seven full-shell and two partial-shell devices had both junctions fully tunable without obvious QD features. %One device is not part of this work, because it revealed quasiparticle poisoning \cite{Albrecht_poisoning_PRL}.

%\subsubsection*{Measurements}
%The differential conductance, $dI/dV$, was measured via a lock-in technique in a dilution refrigerator with a base temperature of $20 \, \si{mK}$.
\subsection*{Theory model}
Full-shell nanowires were modeled as an axially symmetric, internal InAs semiconducting core of radius $R=60$nm, covered by a 25nm-thick epitaxial Al superconducting shell. The electron subbands in the core are classified by their total angular momenta and remain decoupled in our model. The shell induces radial band bending and increased carrier density as one approaches the core/shell interface at $r=R$. Additionally, the shell induces superconducting correlations on core states by the proximity effect. The induced pairing $\Delta$ within lobe $n$ acquires the same fluxoid number $n$ of the parent Al order parameter, $\Delta\sim e^{in\phi}$. The absolute strength of the induced pairing depends on interface transparency, and is assumed here to be the most relevant difference between samples.
The island is coupled to normal leads through $\sim 100$nm-long potential barriers that can be tuned from tunneling to transparent regimes. This model allows us to simulate both the local density of states (LDOS) at any point of the island in a closed/closed barrier configuration, and also the $dI/dV$ transport for closed/open islands. Both calculations are performed within a non-interacting electron approximation using the Quantica.jl package \cite{Quantica:Z21}. Further details on the model are provided in the SI.

\subsection*{Acknowledgment} 
The authors thank P. Krogstrup for providing us with the NW materials. Furthermore, we thank A. Higginbotham, E. J. H. Lee, C. Marcus and S. Vaitiek{\.e}nas for helpful discussions. 

{\textbf{Funding}}. This research was supported by the Scientific Service Units of ISTA through resources provided by the MIBA Machine Shop and the nanofabrication facility; the NOMIS Foundation; the Spanish Ministry of Economy and Competitiveness through Grants FIS2017-84860-R, PCI2018-093026 (FlagERA Topograph) and PGC2018-097018-BI00 (AEI/FEDER, EU), the Comunidad de Madrid
through Grant S2018/NMT-4511 (NMAT2D-CM) and the CSIC Research Platform on Quantum Technologies PTI-001. ICN2 also acknowledges funding from Generalitat de Catalunya 2017 SGR 327. ICN2 is supported by the Severo Ochoa program from Spanish MINECO (Grant No. SEV-2017-0706) and is funded by the CERCA Programme / Generalitat de Catalunya. Part of the present work has been performed in the framework of Universitat Autònoma de Barcelona Materials Science PhD program. Authors acknowledge the use of instrumentation as well as the technical advice provided by the National Facility ELECMI ICTS, node "Laboratorio de Microscopías Avanzadas" at University of Zaragoza.  This project has received funding from the European Union’s Horizon 2020 research and innovation programme under grant agreement No 823717 – ESTEEM3. M.B. acknowledges support from SUR Generalitat de Catalunya and the EU Social Fund; project ref. 2020 FI 00103.%the European Union’s Horizon 2020 research and innovation program under the Marie Sklodowska-Curie grant agreement No 844511; the FETOPEN Grant Agreement No. 828948; the European Research Commission through the grant agreement HEMs-DAM No 716655;  the Spanish Ministry of Science and Innovation through Grants PGC2018-097018-B-I00, PCI2018-093026, FIS2016-80434-P (AEI/FEDER, EU), RYC-2011-09345 (Ram\'on y Cajal Programme), and the Mar\'ia de Maeztu Programme for Units of Excellence in R\&D (CEX2018-000805-M); 
%the CSIC Research Platform on Quantum Technologies PTI-001. 

{\textbf{Author contribution}}. M.V., M.Bor. and G.K. designed the experiment. M.V. and M.Bor. fabricated the devices, performed the measurements and analyzed the data under the supervision of G.K.. A. H. and G. K. performed pre-characterization measurements on island devices. S.M.-S., M.Bot. and J.A. performed the HAADF STEM and EELS measurements. E.P., R.A. and P.S-J. provided theory support during and after the
measurements, and developed the theoretical framework and models
to analyze the experiment.
E.P and P.S-J. performed the numerical simulations. M. V, M. Bor., E. P., R.A., P.S-J. and G.K. contributed to discussions and the preparation of the manuscript with input from the rest of authors.

{\textbf{Competing interests}}: The authors declare no competing interests. 

{\textbf{Data and Materials Availability}}:All experimental data included in this work and related to this work but not explicitly shown in the paper will be available via the IST Austria repository~[\url{https://doi.org/10.15479/AT:ISTA:12102}]. Code used for the data analysis, microscopy analysis data and the codes used for the numerical simulation can also be found at the IST Austria repository~[\url{https://doi.org/10.15479/AT:ISTA:12102}].

\newpage

%\textbf{Observations and questions:}

%\EditMV{\textit{Previous results reported in the literature were not in the weak tunneling regime [refs] as already pointed out theoretically [refs some Lutchyn, Flensberg and Glazman papers] {I do not know where we should write this sentence...}.}}

%\EditMV{\textit{ we have two mention that increasing transmission we increase the charging energy [ref. Flensberg]}}

%\EditMV{We should mention dynamical coulomb decoupling, maybe in the figure caption... }

%\textit{\begin{itemize}
%    \item We need theory support for the enhancement of subgap conductance at zero bias. Is something typical of full-shell or it can be extended to half-shell?
 %   \item If we have normal DOS, we cannot have MAjoranas(?!) If there is no gap, no Majorana? 
  %  \item we need simulation also of the island experiment in both 2e, 1e with Majorana and 1e for crappy superconductor. (this we can do it, maybe, this depends on the time)
 %   \item \EditMB{there is this paper: https://arxiv.org/pdf/1606.06756.pdf that tells that Majorana peaks should be much broader than SET peaks in a high transmission limit. However, comparing destructive to the 1st lobe, we could see that it's not the case. Maybe we could be qualitative on this point and support ourselves? What about non-Majorana but still a single sub-gap state?}
  %  \item adding full-shell transmon paper along with gapless idea.
  %  \item in the master equation formalism, what happens if we have gapless  + state(s) inside the gap?
%\end{itemize}}

\bibliography{references.bib}

\end{document}